\documentclass[aps,prb,groupedaddress,showpacs,twocolumn]{revtex4}
\usepackage{graphicx}
\usepackage{amsmath}
\usepackage{amssymb}
\usepackage{bbm}
\usepackage{xspace}
\usepackage{color}
\usepackage{footnote}
\usepackage{comment}
\newcommand{\mytr}{\text{tr}}

\definecolor{orange}{RGB}{252,77,6}
\definecolor{brown}{RGB}{200,127,50}
\definecolor{green1}{RGB}{00,100,00}
\definecolor{green2}{RGB}{00,150,00}
\definecolor{green3}{RGB}{00,200,00}
\definecolor{green4}{RGB}{00,250,00}
\definecolor{gray}{RGB}{150,150,150}

\newcommand{\intP}[2]{\int\limits_{#1}^{#2}\hspace{-13pt}\mathcal{P}\,}
\newcommand{\myincludegraphics}[2][]{\includegraphics[#1]{#2}}

\newcommand{\fig}[1]{Fig.\thinspace{}\ref{#1}}

\newcommand{\Fig}[1]{Fig.\thinspace{}\ref{#1}}
\newcommand{\eq}[1]{Eq.\thinspace{}(\ref{#1})}

\newcommand{\se}{Sec.\@\xspace}

\newcommand{\app}{App.\@\xspace}

\newcommand{\etal}[0]{\textit{et al.}}
\newcommand{\tcite}[1]{Ref.~\onlinecite{#1}}

\newcommand{\uu}{1\hspace{-3pt}1}

\def\bra#1{\mathinner{\langle{#1}|}}
\def\ket#1{\mathinner{|{#1}\rangle}}

\newcommand{\nag}{{\phantom{\dag}}}

\newcommand{\tab}[1]{Tab.\thinspace{}\ref{#1}}

\usepackage{hyperref}

\begin{document}

% Use the \preprint command to place your local institutional report
% number in the upper righthand corner of the title page in preprint mode.
% Multiple \preprint commands are allowed.,
% Use the 'preprintnumbers' class option to override journal defaults
% to display numbers if necessary
%\preprint{}

%Title of paper
\title{Master equation based steady-state cluster perturbation theory}

% repeat the \author .. \affiliation  etc. as needed
% \email, \thanks, \homepage, \altaffiliation all apply to the current
% author. Explanatory text should go in the []'s, actual e-mail
% address or url should go in the {}'s for \email and \homepage.
% Please use the appropriate macro foreach each type of information

% \affiliation command applies to all authors since the last
% \affiliation command. The \affiliation command should follow the
% other information
% \affiliation can be followed by \email, \homepage, \thanks as well.
\author{Martin Nuss}
\email[]{martin.nuss@tugraz.at}
\affiliation{Institute of Theoretical and Computational Physics, Graz University of Technology, 8010 Graz, Austria}
\author{Gerhard Dorn}
\affiliation{Institute of Theoretical and Computational Physics, Graz University of Technology, 8010 Graz, Austria}
\author{Antonius Dorda}
\affiliation{Institute of Theoretical and Computational Physics, Graz University of Technology, 8010 Graz, Austria}
\author{Wolfgang von der Linden}
\affiliation{Institute of Theoretical and Computational Physics, Graz University of Technology, 8010 Graz, Austria}
\author{Enrico Arrigoni}
\affiliation{Institute of Theoretical and Computational Physics, Graz University of Technology, 8010 Graz, Austria}

\date{\today}

\begin{abstract}
A simple and efficient approximation scheme to study electronic transport characteristics of strongly correlated nano devices, molecular junctions or heterostructures out of equilibrium is provided by steady-state cluster perturbation theory. In this work, we improve the starting point of this perturbative, nonequilibrium Green's function based method. Specifically, we employ an improved unperturbed (so-called reference) state $\hat{\rho}^S$, constructed as the steady-state of a quantum master equation within the Born-Markov approximation. This resulting hybrid method inherits beneficial aspects of both, the quantum master equation as well as the nonequilibrium Green's function technique. We benchmark the new scheme on two experimentally relevant systems in the single-electron transistor regime: An electron-electron interaction based quantum diode and a triple quantum dot ring junction, which both feature negative differential conductance. The results of the new method improve significantly with respect to the plain quantum master equation treatment at modest additional computational cost.
\end{abstract}

% insert suggested PACS numbers in braces on next line
%71.27+a strongly correlated systems
%71.10.-w theories and models of condensed matter
%71.15.-m condensed matter calculation methods
%72.15.Qm Kondo Effect
%71.55.Ak local magnetic moments in metals
%73.63.Kv       Quantum dots
%73.23.-b       Electronic transport in mesoscopic systems
%72.10.Fk Scattering by point defects, dislocations, surfaces, and other imperfections (including Kondo effect)
%71.15.-m condensed matter calculation methods
%72.15.Qm Kondo Effect
%71.10.-w theories and models of condensed matter
%73.21.La Quantum dots
%75.78.-n Magnetization dynamics
% insert suggested keywords - APS authors don't need to do this
%\keywords{}

%\maketitle must follow title, authors, abstract, \pacs, and \keywords
\maketitle

\section{Introduction}\label{sec:Introduction}
%% GENERAL PROBLEM
Electronic transport in the realm of molecular scale junctions and devices has become a subject of intense study in recent years.~\cite{cu.fa.05, ni.ra.03, ag.ye.03, cu.sc.10, na.bl.09, vent.08, fe.go.09} Nowadays the controlled assembly of structures~\cite{gr.ma.07} via electro migration,~\cite{pa.pa.02, li.sh.02, ku.da.03, yu.ke.05, ch.be.06, po.os.06, he.gr.06, os.on.07, da.ku.08} the contacting in mechanical break-junction setups,~\cite{sm.no.02, ch.pa.05,lo.we.07, ki.ta.08} electronic gating~\cite{gi.be.00, ch.pa.05, da.ku.08} and measurement via scanning tunnelling microscopy~\cite{xi.xu.04, re.me.05, ve.kl.06, ko.am.12} have become established tools, ultimately opening routes from elementary understanding to device engineering. Prompted by these formidable advances in experimental techniques, the characterization of transport through e.g. molecules bound by anchor groups to metal electrodes,~\cite{re.zh.97, lo.we.98, ki.ta.08} heterostructures~\cite{ro.ha.08, ga.ka.09} or nano structures on two-
dimensional substrates~\cite{mo.dr.09, er.we.09, to.jo.07,ga.st.06,ro.ha.08, ga.sa.09, au.pa.10} has become feasible. These constitute the foundation for future applications in electronic devices based on single electron tunnelling,~\cite{ko.ma.97} quantum interference effects,~\cite{barf.05, be.da.08, da.be.09,ca.st.06,ga.so.07, qi.li.08, ke.ya.08} spin control~\cite{do.be.09, do.be.10} or even quantum many-body effects~\cite{ro.fl.08, pa.pa.02, li.sh.02, yu.ke.05, lo.ke.05} like Kondo~\cite{hews.97} behaviour.~\cite{go.go.98, fr.ha.02, le.sc.05, kr.sh.11, kr.sh.12}

%% GENERAL METHODS
Typically the electronic transport through such devices is significantly influenced by electronic correlation effects, which may become large due to the reduced effective dimensionality and/or confined geometries. This is reflected, for instance, in major discrepancies between experimental and theoretical current-voltage characteristics obtained with (uncorrelated) nonequilibrium Green's function~\cite{ha.ja.96,my.st.09,wa.ag.13,ku.vo.07} calculations based on \textit{ab-initio} density functional theory states.~\cite{de.gr.04, st.kr.08, cu.fa.05,ch.th.12,st.th.11} The inclusion of many-body effects in the theoretical description of fermionic systems out of equilibrium~\cite{ry.gu.09,rich.99,ha.ja.96, datt.05} is challenging and an active area of current research.~\cite{an.me.10,scho.09,ro.pa.05,ha.ke.09,ec.he.10,re.ja.15,ao.ts.14,scho.11,ande.08} Suitable approximations need to be devised in order to solve a finite strongly correlated quantum many-body problem out of equilibrium coupled to an infinite environment. Typically, the nonequilibrium setup consists of a correlated central region (system) attached to two leads (environment).

%% QME
A well-established method for treating such open quantum systems is by means of quantum master equations (Qme).~\cite{br.pe.02, carm.93, carm.10, scha.12, scha.14, ta.du.98} Herein, the environment-degrees of freedom are integrated out and usually incorporated in a perturbative manner. The Qme approach allows a detailed investigation of transport phenomena~\cite{do.be.09, do.be.10} and recent self-consistent extensions attempt to cure some of its long-standing limitations.~\cite{ji.li.14}

%% CPT
In the framework of nonequilibrium Green's functions (NEGF) various schemes exist to approximately calculate the electronic self-energy of the correlated region, see e.g. \tcite{me.wi.92,sc.sc.94, ro.pa.05, ec.ha.09, ge.pr.07, ja.me.07, fr.tu.06}. In cluster approaches, such as cluster perturbation theory (CPT) and its improvement, the variational cluster approach (VCA),~\cite{po.ai.03} the whole system is partitioned into parts which can be treated exactly and determine the self-energy. Originally devised for strongly correlated systems in equilibrium,~\cite{gr.va.93,se.pe.00} both approaches have recently been extended to nonequilibrium situations in the time dependent case~\cite{ba.po.11, ho.ec.13} as well as in the steady-state.~\cite{kn.li.11, ho.ec.13} In previous work we applied the steady-state CPT (stsCPT) to obtain transport characteristics of heterostructures,~\cite{kn.li.11} quantum dots~\cite{nu.ar.12a, nu.he.12, nu.ar.12b} and molecular 
junctions~\cite{nu.li.13, kn.ar.13} and obtained good results even in the challenging Kondo regime.~\cite{hews.97,nu.ar.12a,
nu.he.12}

%% SPECIFIC PROBLEM
A key issue in the CPT approach is to identify an appropriate many-body state for the disconnected correlated cluster in the central region, as a starting point of  perturbation theory, the so-called reference state. Up to now, a common choice in stsCPT is to use an equilibrium state at some  temperature $T_S$ (often $T_S=0$) and chemical potential $\mu_S$ in-between the values of the leads. Such an ad-hoc choice is clearly unsatisfactory. Furthermore, it fails to describe certain quantum interference effects in transport phenomena as for example so-called current blocking effects.~\cite{nu.li.13,do.be.09, do.be.10}

%% ANSWER
The purpose of the present work is to improve on stsCPT by constructing a consistent and conceptually more appropriate reference state, given by the steady-state reduced many-body density matrix $\hat{\rho}^S$ obtained from a Qme in the Born-Markov approximation. Within this quantum master equation based stsCPT (meCPT), the ambiguity in defining $\mu_S$ and $T_S$ for the central region is resolved. The equilibrium case, in which  $\mu_S$ and $T_S$ coincide with those of the environment, is automatically included. In contrast to standard Qme approaches, lead induced level-broadening effects are accounted for and the noninteracting limit is reproduced exactly, as in the original stsCPT. In addition, meCPT is able to capture the previously mentioned current blocking effects, as shown below.

%% NEGF + Qme hybrids
Other NEGF/Qme hybrid methods exist in the literature.\cite{scho.09,dz.ko.12,dz.ko.15} For instance, in a recent work~\cite{ar.kn.13,do.nu.14} we have proposed a so-called auxiliary master equation (AME) approach, whereby a Lindblad equation is introduced which models the leads by a small number of bath sites plus Markovian environments. The AME is suited to address steady-state properties of single impurity problems as encountered in the framework of nonequilibrium dynamical mean field theory.~\cite{ar.kn.13,ge.ko.96,me.vo.89,sc.mo.02u,fr.tu.06,ao.ts.14} In contrast, the meCPT presented in this work is more appropriate to treat non-local self-energy effects which cannot be captured by single-site DMFT.

%% OUTLINE
This paper is organized as follows: After defining the model Hamiltonian in \se~\ref{sec:model}, the meCPT is introduced in detail in \se~\ref{sec:meCPT}. We present results obtained with the improved method for two experimentally realizable devices: i) In \se~\ref{ssec:MBlocking}, an electron-electron interaction based quantum diode, ii) and in \se~\ref{ssec:QiBlocking}, a triple quantum dot ring junction which both feature negative differential conductance (NDC).

For ring systems, extensive Qme results and an explanation of the NDC in terms of quantum interference mediated blocking are available in \tcite{do.be.09, do.be.10}.

\section{Model}\label{sec:model}
\begin{subequations}
%% GENERAL
We consider a model of spin-$\frac{1}{2}$ fermions, having in mind the electronic degrees of freedom of a contacted nano structure, heterostructure or a molecular junction. The Hamiltonian consists of three parts: 
\begin{align}
\hat{\mathcal{H}} &= \hat{\mathcal{H}}^{S} + \hat{\mathcal{H}}^{E} + \hat{\mathcal{H}}^{SE}\,\mbox{.}
\label{eq:H}
\end{align}
%% SYSTEM
i) The ``system'' $\hat{\mathcal{H}}^{S}$ represents the interacting central region i.e. the nano device or molecule consisting of single-particle as well as interaction many-body terms. It is described by electronic annihilation/creation operators $f^\nag_{i\sigma}/f^\dag_{i\sigma}$ at site $i=[1,\ldots,N_S]$ where $N_S$ is typically small and spin $\sigma=\{\uparrow,\downarrow\}$.~\cite{ne.or.98} We will specify the particular form of $\hat{\mathcal{H}}^{S}$ in the respective results section.
%% ENVIRONMENT
ii) The ``environment'' Hamiltonian $\hat{\mathcal{H}}^{E}$ describes the two noninteracting electronic leads 
\begin{align}                                                                                                                                                                                                                                                                                                                                                                                                  
\hat{\mathcal{H}}^{E}&=\sum\limits_{\lambda=1}^2\sum\limits_{k\sigma}\epsilon^\nag_{\lambda k \sigma} c_{\lambda k \sigma}^\dag c_{\lambda k \sigma}^\nag\,\mbox{,}
\label{eq:He}
\end{align}
where $c^\nag_{\lambda k \sigma}/c^\dag_{\lambda k \sigma}$ denote the fermion operators of the infinite size lead $\lambda$ with energies $\epsilon_{\lambda k \sigma}$ and electronic density of states (DOS) $\rho_{\lambda \sigma}(\omega)=\frac{1}{N_\lambda}\sum\limits_k \delta\left(\omega-\epsilon_{\lambda k \sigma}\right)$ where $N_\lambda\to\infty$ are the number of levels in the leads. The disconnected leads are held at constant temperatures $T_\lambda$ and chemical potentials $\mu_\lambda$ so that the particles obey the Fermi-Dirac distribution $p^{\text{FD}}_\lambda(\omega,T_\lambda,\mu_\lambda)$.~\cite{ne.or.98, footnote2} 
%% COUPLING
iii) Finally the system and the environment are coupled by the single-particle hopping
\begin{align}                                                                                                                                                                                                                                                                                                                                                                                                  
\hat{\mathcal{H}}^{SE} &=\sum\limits_{\lambda=1}^{2} \sum\limits_{ik\sigma}\left(t'_{\lambda i k \sigma} f_{i\sigma}^\dag c_{\lambda k \sigma}^\nag\right) + h.c. \,\mbox{.}
\label{eq:Hc}
\end{align}
\end{subequations}

\section{Master Equation based Cluster Perturbation Theory}\label{sec:meCPT}
%% INTRODUCE PROBLEM
Our goal is to obtain the steady-state transport characteristics of the Hamiltonian $\hat{\mathcal{H}}$, \eq{eq:H} in a nonequilibrium situation induced by environment parameters like a bias voltage $V_B$ or temperature gradient $\Delta T$. The important step consists in evaluating the steady-state single-particle Green's function in Keldysh space $\widetilde{G}$ in the well established Keldysh-Schwinger nonequilibrium Green's function formalism.~\cite{schw.61, feyn.63, keld.65} In general $\hat{\mathcal{H}}$ is both interacting and of infinite spatial extent. Therefore explicit evaluation of $\widetilde{G}$ is prohibitive in all but the most simple cases which motivates the introduction of approximate schemes.

%% INTRODUCE CHALLENGE
One such scheme is CPT,~\cite{gr.va.93,se.pe.00} in which one performs an expansion in a 'small' single-particle perturbation, for example the system-environment coupling $\hat{\mathcal{H}}^{SE}$ of \eq{eq:Hc}. The unperturbed Hamiltonian $\hat{\mathcal{H}}^{S}+\hat{\mathcal{H}}^{E}$ can be solved exactly. While in the noninteracting case CPT becomes exact, results obtained in the presence of interaction are approximate and depend on the reference state for the unperturbed system. A common practice within stsCPT~\cite{nu.ar.12a, nu.he.12, nu.ar.12b, nu.li.13, kn.ar.13, ho.ec.13} is to use a pure state given by the equilibrium ground state $\ket{\Psi_0}_S$ of the disconnected interacting system Hamiltonian $\hat{\mathcal{H}}^S$. In a nonequilibrium situation, this is still ambiguous, as it depends on an arbitrary choice of the chemical potential $\mu_S$ and/or temperature $T_S$ for the interacting finite system. 

%% NEW REFERENCE STATE
The goal of this work is to provide an unambiguous and conceptually more rigorous criterion for the choice of the reference state for the interacting central region. Ideally, the reference state is selected such that it resembles best the situation of the coupled system, i.e. for the full Hamiltonian, \eq{eq:H} in the steady-state. An appropriate choice in equilibrium is to use the grand-canonical density operator~\cite{ne.or.98} $\hat{\rho}^S_{gc}$ as reference state. In this case, $T_S$ and $\mu_S$ are uniquely determined by the equilibrium situation. Equivalently, $\hat{\rho}^S_{gc}$ is given by the steady-state solution of a Qme in the Born-Markov approximation (see \se~\ref{ssec:BMsme}), when coupling the system to one thermal environment. From this viewpoint a natural extension to the nonequilibrium situation is to make use of a Qme as well in order to obtain a consistent reference state, given then by the steady-state reduced density operator of the system $\hat{\rho}^S$. In this work, a second order Born-Markov Qme is employed, which yields the correct zeroth order reduced density operator $\hat{\rho}^S$ (adjusted to $\hat{\mathcal{H}}^{SE}$).~\cite{fl.cu.11,mo.mi.08} Subsequently, $\hat{\mathcal{H}}^{SE}$ is included within the CPT approximation,~\cite{gr.va.93,se.pe.00} in order to obtain improved results for the Green's function and in turn for the transport observables.

In summary, the meCPT method consists of the following three main steps, analogous to a standard CPT treatment:
\begin{enumerate}
 \item Decompose the whole system into a small interacting central region (system) and noninteracting leads of infinite size (environment), see $\hat{\mathcal{H}}^{S}$ and $\hat{\mathcal{H}}^{E}$ in \eq{eq:H}. 
 \item The \emph{new step introduced in this work} is to solve a Qme for the system in order to obtain the reduced density operator $\hat{\rho}^S$, which serves as a reference state to calculate the cluster (retarded) Green's function~\cite{footnote4}
 \begin{align}
 g^R_{ij\sigma}(\tau)&=-i\theta(\tau)\text{tr}\left\{\hat{\rho}^S\left[f^\nag_{i\sigma}(\tau),f^\dag_{j\sigma}\right]_{+}\right\}\,\mbox{.}
\label{eq:gr}
\end{align}
\item Reintroduce the system-environment coupling $\hat{\mathcal{H}}^{SE}$ perturbatively, see \se~\ref{ssec:stsCPT} and \eq{eq:CPT}, to determine the Green's function of the coupled system.
\end{enumerate}

\subsection{Steady-state cluster perturbation theory}\label{ssec:stsCPT}
%% CPT INTRO
Here we briefly recall the main, well-established CPT concepts and equations, as this is the starting point for the formalism presented in this work. 
For an in depth discussion of CPT~\cite{sene.09} and its nonequilibrium extension we refer the reader to the literature.~\cite{ba.po.11,kn.li.11, nu.he.12, nu.li.13} 

%% KELDYSH GF
The central element of stsCPT is the steady-state single-particle Green's function in Keldysh space~\cite{ra.sm.86}
\begin{equation}
 \widetilde{G} = \begin{pmatrix} G^R & G^K \\ 0 & G^A \end{pmatrix}\,,
\end{equation}
where $R$ denotes the retarded, $A$ the advanced, and $K$ the Keldysh component. In the present formalism, $G^{R/A/K}$ become matrices in the space of cluster sites and depend on one energy variable $\omega$ since time translational invariance applies in the steady-state. 

%% QUICKLY RECAPITULATE CPT
As explained above, in order to compute $\widetilde{G}(\omega)$ within stsCPT one partitions $\hat{\mathcal{H}}$, \eq{eq:H} in real space, into individually exactly solvable parts, in this case, the system $\hat{\mathcal{H}}^S$ and the environment $\hat{\mathcal{H}}^E$, which leaves the coupling Hamiltonian $\mathcal{H}^{SE}$ as a perturbation. The single-particle Green's function of the disconnected Hamiltonian is denoted by $\widetilde{g}(\omega)$, which obviously does not mix the disconnected regions. For the noninteracting environment, the respective block entries of $\widetilde{g}(\omega)$ are available analytically.~\cite{econ.10, nu.ar.12a} For the interacting part the respective entries of $\widetilde{g}(\omega)$ are calculated via the Lehmann representation with respect to the reference state. This can be computed e.g. based on the Band Lanczos method.~\cite{ba.de.87, press.07, footnote3} 

The full steady-state Green's function in the CPT approximation is found by reintroducing the inter-cluster coupling perturbatively
\begin{align}
 \widetilde{G}(\omega)^{-1}&=\widetilde{g}(\omega)^{-1}-\widetilde{M}\,\mbox{;} 
\quad
M^R=M^A=M \,,\hspace{0.5em} M^K=0\,\mbox{,}
\label{eq:CPT}
\end{align}
where we denote by the matrix $M$ the single-particle Wannier representation of $\hat{\mathcal{H}}^{SE}$. CPT is equivalent to using the self-energy $\widetilde{\Sigma}$ of the disconnected Hamiltonian as an approximation to the full self-energy.  Therefore, the quality of the approximation can in principle be systematically improved by adding more and more sites of the leads to the central cluster. However, in doing so the complexity for the exact solution of the central cluster grows exponentially. Independent of the reference state, this scheme becomes exact in the noninteracting limit.

\subsection{Born-Markov equation for the reference state}\label{ssec:BMsme}
%% REFERENCE STATE 
%% INTRO
In the following we outline how to obtain the reference state $\hat{\rho}^S$ by using a Born-Markov-secular (BMsme), or more generally a Born-Markov master equation (BMme).~\cite{br.pe.02, carm.93, carm.10, scha.12, scha.14, ta.du.98} Although this approach is standard, for completeness we present here the main aspects and notation. We loosely follow the treatment of \tcite{scha.12, scha.14, da.be.09}.

The real time $\tau$ evolution of the full many-body density matrix $\hat{\rho}$ is given by the von-Neumann equation $\dot{\hat{\rho}}=-i\left[\hat{\mathcal{H}},\hat{\rho}\right]_-$.~\cite{br.pe.02} Typically the large size of the Hilbert space of $\hat{\mathcal{H}}$ prohibits the full solution in the interacting case. One thus considers the weak coupling limit $|\hat{\mathcal{H}}^{SE}|\ll|\hat{\mathcal{H}}^{E}|$ and performs a perturbation theory in terms of $|\hat{\mathcal{H}}^{SE}|$.~\cite{ta.du.98, footnote1} 

%% VONNEUMANN TO BMSME
In the usual way one obtains an equation for the reduced many-body density matrix of the system $\hat{\rho}^S(\tau) = \mytr_E\left\{ \hat{\rho} \right\}$ by working in the interaction picture $\hat{\rho}_I(\tau) = e^{+i(\hat{\mathcal{H}}^S+\hat{\mathcal{H}}^E) \tau}\hat{\rho}(0)e^{-i(\hat{\mathcal{H}}^S+\hat{\mathcal{H}}^E)\tau}$ with respect to the coupling Hamiltonian, \eq{eq:Hc}. One then performs three standard approximations: 
i) Within the Born approximation, valid to lowest order in $|\hat{\mathcal{H}}^{SE}|$, the density matrix is factorized $\hat{\rho}_I(\tau)\approx \hat{\rho}^S_I(\tau)\otimes{\hat{\rho}}^{E}_I $. Furthermore, the environment ${\hat{\rho}}^E_I$ is assumed to be
so large that it is not affected by $|\hat{\mathcal{H}}^{SE}|$ and thus independent of time.
ii) The Markov approximation implies a memory-less environment, that is, the system density matrix varies much slower in time than the decay time of the environment correlation functions $C_{\alpha\beta}(\tau)$. Upon transforming back to the 
Schr\"odinger picture this yields the BMme, which is time-local, preserves trace and hermiticity, and depends on constant coefficients. 
iii) To obtain an equation of Lindblad form which also preserves positivity one typically employs the secular approximation, which averages over fast oscillating terms, yielding the BMsme.~\cite{scha.12,whit.08,ya.sh.00}

%% BMSME
The system-environment coupling  can be quite generally written in the form $\hat{\mathcal{H}}^{SE}=\sum\limits_\alpha\hat{S}_\alpha\otimes \hat{E}_\alpha$, with $\hat{S}_\alpha = \hat{S}_\alpha^\dag$ and $\hat{E}_\alpha = \hat{E}_\alpha^\dag$. This hermitian form is convenient for further treatment.The tensor product form can be achieved even for fermions by a Jordan-Wigner transformation,~\cite{scha.14} see \app~\ref{app:tensorProduct}. For our coupling Hamiltonian, \eq{eq:Hc} and particle number conserving systems, the coupling operators take the form
\begin{align}
\label{eq:couplingOperators}\hat{S}_{1i\sigma} &= \frac{1}{\sqrt{2}}(f_{i\sigma} + f_{i\sigma}^\dag), 
\hat{E}_{1\lambda i\sigma} = \frac{1}{\sqrt{2}}(c_{\lambda i\sigma} + c_{\lambda i\sigma}^\dag)\\
\nonumber \hat{S}_{2i\sigma} &= \frac{i}{\sqrt{2}}(f_{i\sigma} - f_{i\sigma}^\dag), 
\hat{E}_{2\lambda i\sigma} = \frac{i}{\sqrt{2}}(c_{\lambda i\sigma} - c_{\lambda i\sigma}^\dag)\,\mbox{.}
\end{align}

In the energy eigenbasis of the system Hamiltonian $\hat{\mathcal{H}}^S\ket{a} = \omega_a\ket{a}$, the BMme 
in the Schr\"odinger representation reads~\cite{footnote4}
\begin{align}
\label{eq:BMsme}
 \dot{\hat{\rho}}^{S}(\tau)&=-i\left[\hat{\mathcal{H}}^S+\hat{\mathcal{H}}^{LS},\hat{\rho}^S(\tau)\right]_- + \sum\limits_{abcd} \Xi_{ab,cd} \nonumber\\
 & \left(\ket{a}\bra{b}\hat{\rho}^{S}(\tau)\ket{d}\bra{c}-\frac{1}{2}\bigg[\ket{d}\bra{c}\ket{a}\bra{b},\hat{\rho}^{S}(\tau)\bigg]_{+}\right)\,\mbox{,}
\end{align}
with
\begin{align}
\label{eq:Xi}\Xi_{ab,cd} &= \sum\limits_{\alpha\beta} \xi_{\alpha\beta}(\omega_{ba},\omega_{dc}) \bra{a}\hat{S}_\beta\ket{b} \bra{c}\hat{S}_\alpha\ket{d}^*\,\mbox{,}
\end{align}
where $\omega_{ba} = \omega_{b}-\omega_{a}$.
The Lamb-shift Hamiltonian $\hat{\mathcal{H}}^{LS}$ and the environment functions $\xi_{\alpha\beta}(\omega_1,\omega_2)$ are defined in \app~\ref{app:BMdetails}. When employing the secular approximation, the terms in the BMsme simplify and in \eq{eq:Xi} one can replace $\xi_{\alpha\beta}(\omega_{ba},\omega_{dc}) \rightarrow \xi_{\alpha\beta}(\omega_b-\omega_a)\delta_{\omega_b-\omega_a,\omega_d-\omega_c}$. Due to the secular approximation the BMsme can only lead to interference between degenerate states. The more general BMme also couples non-degenerate states at the cost of loosing the Lindblad structure of the Qme, see \se~\ref{ssec:Vg3} and \tcite{da.be.09}.

\subsubsection*{Single-particle Green's function}\label{sssec:GfFromMbDm}
As discussed above, for meCPT, the Green's function $\widetilde{g}(\omega)$ of the isolated system is evaluated from the reference state $\hat{\rho}^S$. The retarded component \eq{eq:gr} takes the explicit form
\begin{align}
\label{eq:grl}
 g^R_{ij(\sigma)}(\omega) &=\sum\limits_{a b c} \rho^S_{a b}\times\\
\nonumber&\bigg( \frac{\bra{b} f_{i\sigma}^\nag \ket{c}\bra{c} f_{j\sigma}^\dag \ket{a}}{\omega+i0^+-(\omega_c - \omega_b)}+\frac{\bra{b} f_{j\sigma}^\dag  \ket{c}\bra{c} f_{i\sigma}^\nag \ket{a}}{\omega+i0^+-(\omega_a- \omega_c)}
\bigg)\,\mbox{,}
\end{align}
where $i,\,j$ denote indices of system sites.
The advanced component follows from $g^A = \left(g^{R}\right)^\dag$ and the Keldysh component $g^K$ of the finite, unperturbed system  is not relevant for the CPT equation, \eq{eq:CPT}. Once $\widetilde{g}$ is obtained, the full Green's function is again approximately obtained within CPT by \eq{eq:CPT}. Notice that for $U=0$, $\widetilde{G}$ is independent of the reference state, which is why stsCPT, stsVCA as well as meCPT coincide (and become exact) in the noninteracting case.

\subsection{Numerical implementation}\label{ssec:solution}
%% INTRO
From a numerical point of view, the two main steps are to first obtain the reference state $\hat{\rho}^S$ by solving the  Qme and then to evaluate the Green's functions using \eq{eq:grl} and \eq{eq:CPT}. For the solution of the BMme, \eq{eq:BMsme} one needs to carry out the following: i) Full diagonalization of the interacting system Hamiltonian which is done in \emph{LAPACK}, making use of the block structure in $\hat{N}$ and $\hat{S}^z$. ii) Evaluation of the coefficients of the BMme in \eq{eq:BMsme}, which involves coupling matrix elements $\bra{a}\hat{S}_\alpha \ket{b}$ and numerical integration of the bath correlations functions, see \app~\ref{app:BMdetails},\thinspace\ref{sssec:environmentCorrelators}, for which an adaptive Gauss-Kronrod scheme is employed. iii) The steady-state $\hat{\rho}^S$ is finally obtained from the unique eigenvector with eigenvalue zero of \eq{eq:BMsme}, which we determine by a sparse Arnoldi diagonalization. Again, a block structure is related to $\hat{N}$ and $\hat{S}^z$. The 
numerical effort for the exact diagonalization scales with the size of the Hilbert space, and therefore exponentially with the system size $N_S$. In the second major step, the Green's function of the disconnected system is calculated by \eq{eq:grl}. Finally, the meCPT Green's function $\widetilde{G}(\omega)$ is found using \eq{eq:CPT}. We outline how to evaluate observables within meCPT and the Qme in \app~\ref{app:observables}.

\section{Results}\label{sec:Results}
In this section we present results obtained from the meCPT approach. In all calculations, except the ones in \se~\ref{ssec:Vg3}, the secular approximation is applied for the reference state $\hat{\rho}^S$. The main improvements of meCPT with respect to bare BMsme are  i) the inclusion of lead induced broadening effects, ii) the correct $U=0$ limit and iii) a correction for effects missed by an improper treatment of quasi degenerate states in the BMsme (see below). In comparison to the previous ``standard'' stsCPT, meCPT also captures current blocking effects, which are discussed in detail in \tcite{be.da.08} and \tcite{da.be.09} within a Qme treatment.

\subsection{Quantum dot diode}\label{ssec:MBlocking}

\begin{figure}
\myincludegraphics[width=0.4\textwidth]{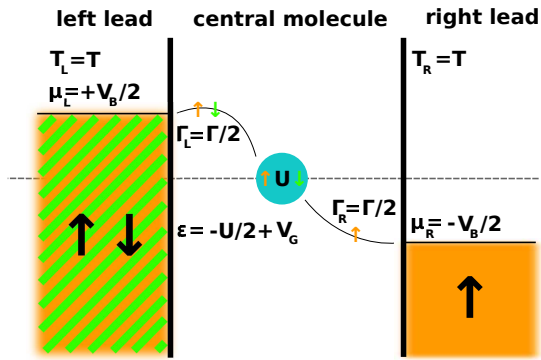}
\caption{(Color online) \emph{Quantum dot diode:}
Schematic representation, see \se~\ref{ssec:MBlocking}. Single quantum dot with Hubbard interaction $U$ and gate voltage $V_G$ (particle-hole symmetric at $V_G=0$), coupled via $\Gamma_{\rm{L/R}}=\frac{\Gamma}{2}$ to a left and right lead. The right lead is fully polarized, i.e. only spin-$\uparrow$ DOS is present. An external bias voltage $V_B$ shifts the chemical potentials by $\mu_{\rm{L/R}}=\pm\frac{V_B}{2}$. The leads are in the wide band limit and at the same temperature $T$.}
\label{fig:fig2}
\end{figure}

%% INTRO
We first discuss a quite simple model system: a quantum diode based on electron-electron interaction effects.
%% DEFINE MODEL
\Fig{fig:fig2} depicts this junction consisting of a single interacting orbital
described by a Hubbard interaction and an on-site term to allow for a gate voltage $V_G$:~\cite{ande.61}
\begin{align*}
 \hat{\mathcal{H}}^S&=U \left(\hat{n}^f_{ \uparrow}-\frac{1}{2}\right) \left(\hat{n}^f_{\downarrow}-\frac{1}{2}\right) +V_G \sum\limits_{\sigma} \hat{n}^f_{\sigma}\,\mbox{,}
\end{align*}
where $\hat{n}^f_{\sigma} = f^\dagger_{\sigma} f_{\sigma}$.
%% DEFINE ENVIRONMENT
The environment \eq{eq:He}, consists of two spin dependent, conducting leads. We model both, the left (L) and the right (R) lead by a flat DOS with  local retarded single-particle Green's function~\cite{econ.10} $g_{L/R}^R(\omega)=-\frac{1}{2 D} \text{ln}\left(\frac{\omega+i0^+-D}{\omega+i0^++D}\right)$, with a half-bandwidth $D$ much larger than all other energy scales in the model, mimicking a wide band limit. We keep both leads at the same temperature $T_L=T_R=T$ and at chemical potentials $\mu_L=-\mu_R=\frac{V_B}{2}$ corresponding to a symmetrically applied bias voltage $V_B$. The right lead is fully spin polarized, i.e. tunnelling of one spin species ($\downarrow$) into the right lead is prohibited while both spin species can tunnel to the left lead.
%% DEFINE COUPLING
The system is coupled to the two leads via a single-particle hopping amplitude $t'$ in $\hat{\mathcal{H}}^{SE}$, \eq{eq:Hc} which results in a lead broadening parameter of $\Gamma_L^\uparrow=\Gamma_L^\downarrow=\Gamma_R^\uparrow=\frac{\Gamma}{2}=\pi|t'|^2\frac{1}{2 D}$, \eq{eq:Gamma}, and $\Gamma_R^\downarrow\equiv 0$. We use $\Gamma$ without an argument for $\Gamma(\omega=0)$ as defined in \eq{eq:Gamma}. For meCPT we use $\mathcal{H}^{SE}$, see \eq{eq:Hc}, as perturbation.

%% HOW TO REALIZE
Such a system could be realized in: i) A ``metal - artificial atom - half-metallic ferromagnet``~\cite{ka.ir.08} nano structure where spin-$\uparrow$ DOS is present at the Fermi energy while the respective spin-$\downarrow$ DOS is zero. ii)
A graphene nano structure~\cite{mo.dr.09, er.we.09} with ferromagnetic cobalt electrodes.~\cite{to.jo.07} iii) A one dimensional optical lattice of ultra cold fermions in a quantum simulator~\cite{bl.da.08} where the hopping of spin-$\downarrow$ particles into the right reservoir is suppressed. For all three systems spin-$\downarrow$ particles cannot reach the right lead, in the first two due to a vanishing DOS, in the third one due to a vanishing tunnelling amplitude.

%% DISCUSS PHYSICS
We consider parameters such that the junction is operated in a single electron transistor (SET) regime,~\cite{ko.ma.97} i.e. temperatures above the Kondo temperature.~\cite{hews.97} In this regime we expect an interaction induced - magnetization mediated blocking due to the fact that the system fills up with spin-$\downarrow$ particles. On the one hand they cannot escape, yielding a vanishing spin-$\downarrow$ current, and on the other hand they suppress the spin-$\uparrow$ occupation, at finite repulsive interaction $U$, resulting also in a vanishing spin-$\uparrow$ current.~\cite{do.be.10}

\begin{figure*}
\myincludegraphics[width=0.95\textwidth]{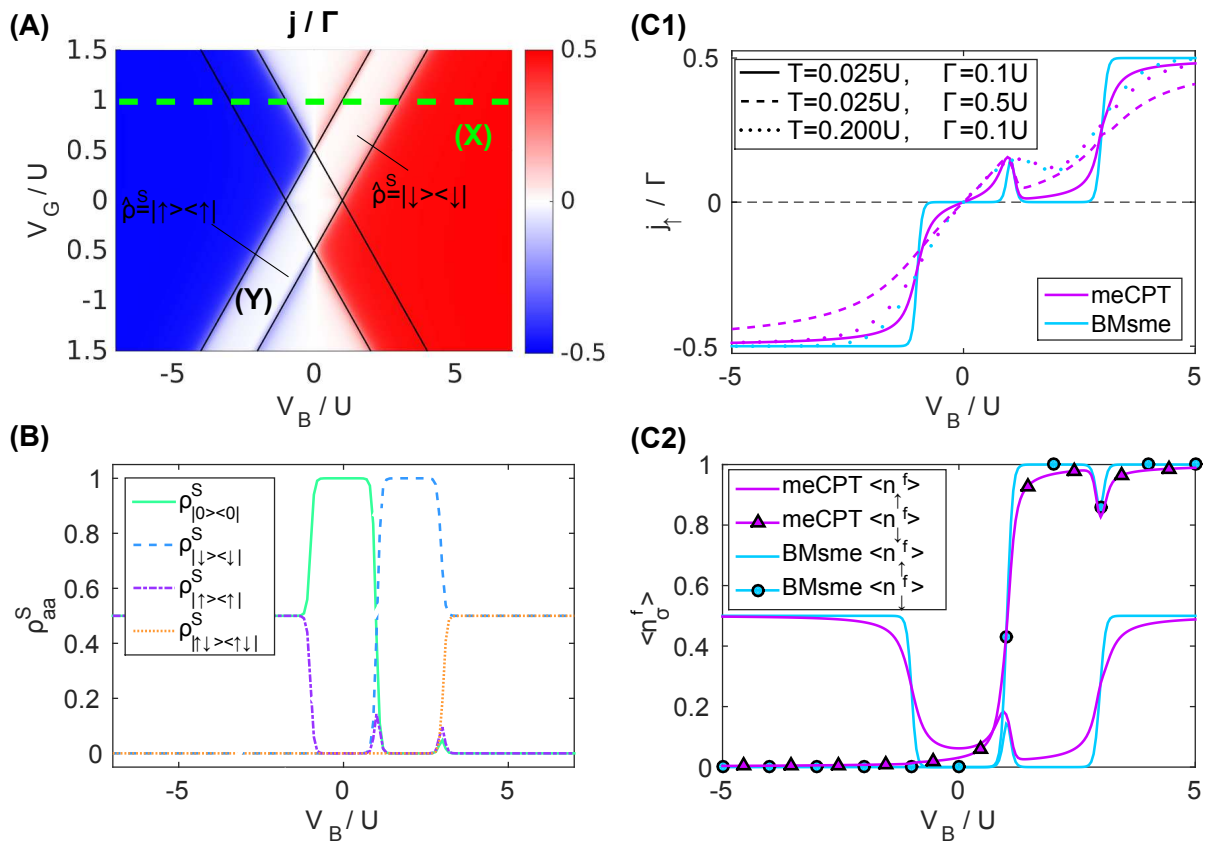}
\caption{(Color online) \emph{Quantum dot diode:}
\textbf{(A)}  Stability diagram, based on the total current $j = \langle j_\uparrow \rangle+\langle j_\downarrow\rangle$ as a function of bias voltage
$V_B$  and gate voltage $V_G$, obtained within
meCPT. Note that $\langle j_\downarrow \rangle\equiv 0$. Results are depicted for
$T=0.025\,U$ and $\Gamma=0.1\,U$. (Y) marks the current blocking region. The green dashed line (X) at $V_G=U$ indicates the parameter regime for the panels (B) and (C).
\textbf{(B)} Diagonal part of the reduced density matrix $\rho^{S}_{aa}$ obtained by BMsme.
\textbf{(C1)} Spin-$\uparrow$ current $ j_\uparrow $ within meCPT compared to BMsme. Solid lines are for the same parameters as line (X) in panel (A). Blue dashed and solid lines for BMsme are indistinguishable. 
\textbf{(C2)} Spin resolved densities $\langle n_\uparrow \rangle$ and $\langle n_\downarrow \rangle$ for the same parameters as in panel (C1), see solid lines in the legend.}
\label{fig:fig3}
\end{figure*}

%% RESULTS: CURRENT
\Fig{fig:fig3} (A) shows the meCPT stability diagram of the interacting system in the $V_B - V_G$ plane. When applying a particle-hole transformation for all particles, leads and system, along with $t'\to-t'$ we easily find the symmetry properties
\begin{align*}
j(-V_{B},-V_{G}) &= -j(V_{B},V_{G})\,\mbox{,}\\[1ex]
\langle n^f_\sigma\rangle(-V_B,-V_G) &= 1-\langle n^f_\sigma \rangle(V_B,V_G)\,\mbox{.}
\end{align*}
From the continuity equation it is clear that only spin-$\uparrow$ steady-state current can flow which limits the maximum current to $\frac{\Gamma}{2}$. The energies $\omega_N$ of the isolated quantum dot can be labelled by the total particle number N and are for $V_{G}=U$ given by $\omega_0=0$, $\omega_1=\frac{1}{2}\,U$ and $\omega_2=2\,U$. This gate voltage corresponds to the dashed line, marked by (X) in \fig{fig:fig3} (A). The corresponding energy differences
$\Delta_{01} =0.5\,U$ between the single-occupied and the empty  dot and $\Delta_{12}=1.5\,U$ between  double-occupied and single-occupied dot are associated with a further transport channel opening  as soon as the bias $V_B$ reaches twice their values. The meCPT result for the current exhibits the well known Coulomb diamond~\cite{ko.ma.97} close to $V_B=0$ and $V_G=0$, where current is hindered because all system energies are far outside the transport window $\pm\frac{V_B}{2}$, see \eq{eq:Wtrans}.  At $V_G=0$ a current sets in at $\frac{V_B}{2}=\pm|\frac{U}{2}|$, i.e. when transport across the system's single particle level becomes allowed. The point, at which the current sets in, shifts with $V_G$ linearly to higher bias voltages. This transition is broadened $\propto \text{max}(\Gamma=0.1\,U, T=0.025\,U)$. However, not only the transport window and possible excitations in the system energies determine the current-voltage characteristics. The particular occupation of the system states may lead to more 
complicated effects, such 
as current blocking.

%% BLOCKING
Our first main result is that in contrast to stsCPT the blocking is correctly reproduced in meCPT. The current blocking is visible in \fig{fig:fig3} (A) in region (Y), see also the detailed data in subplot (C1). It is asymmetric in $V_B$ and therefore responsible for the rectifying behaviour for
$|V_G|>|\frac{U}{2}|$. This feature is easily understood from the
plots of the spin resolved densities in \fig{fig:fig3} (C2). In the
region of interest, for positive $V_B$, $\langle n_\downarrow
\rangle=1$ which hinders spin-$\uparrow$ particles from the left lead
to enter the system, due to the repulsive interaction $U$ and
suppresses the current. For negative $V_B$, the situation is reversed.
A direct computation of the current in the framework of BMsme, see
\app~\ref{app:observables2}, also predicts the blocking, which is
however not the case if we use stsCPT based on the zero temperature
ground state $\ket{\Psi_0}_S$.  
The blocking is evident in \fig{fig:fig3} (B), where we observe that in the blocking regime, the reduced density is $\rho_{S}=\ket{\downarrow}\bra{\downarrow}$. Independent of the value of $U>0$, the blocking sets in at the same values of  $V_B$ in meCPT and BMsme. \fig{fig:fig3} (C1) shows that  within BMsme this regime is entered after a $U$ independent hump in the current while within meCPT the hump is broader and weakly $U$ dependent. The current blocking disappears at a  bias voltage $V_B\propto U$ in both methods. Immediately apparent are the much broader features in meCPT, which leads to a less pronounced effect in contrast to the total blocking predicted by BMsme. In BMsme the broadening parameter $\Gamma$  enters merely as prefactor of the current, and broadening is solely induced by the temperature. This temperature induced broadening is correctly taken into account in both
methods. For  $T>\Gamma$ the latter dominates and the meCPT results are similar to the plain BMsme solution.
A comparison of the three methods is given in \tab{table:table1}. In this simple model the blocking can be captured even by a straight forward steady-state mean-field theory in the Keldysh Green's function with self-consistently determined spin densities or in stsVCA. This is not the case for the more elaborate system studied in the next section.

\begin{table}[ht]
\caption{
Comparison of steady-state cluster perturbation theory (stsCPT), the Born-Markov-secular master equation (BMsme) and the quantum master equation based stsCPT (meCPT) with respect to their ability to capture temperature ($T$) or lead ($\Gamma$) induced level broadening, current blocking and whether the noninteracting limit is fulfilled.}
\centering
\begin{tabular}{c c c c c}
\hline
\hline 
method & $T$-broadening & $\Gamma$-broadening & blocking & $U=0$\\
[0.5ex]
\hline
stsCPT       & yes & yes & no & exact\\
BMsme        & yes & no  & yes& approx.\\
meCPT & yes & yes & yes& exact\\
[1ex]
\hline
\end{tabular}
\label{table:table1}
\end{table}

\subsection{Triple quantum dot}\label{ssec:QiBlocking}
\begin{figure*}
\myincludegraphics[width=0.99\textwidth]{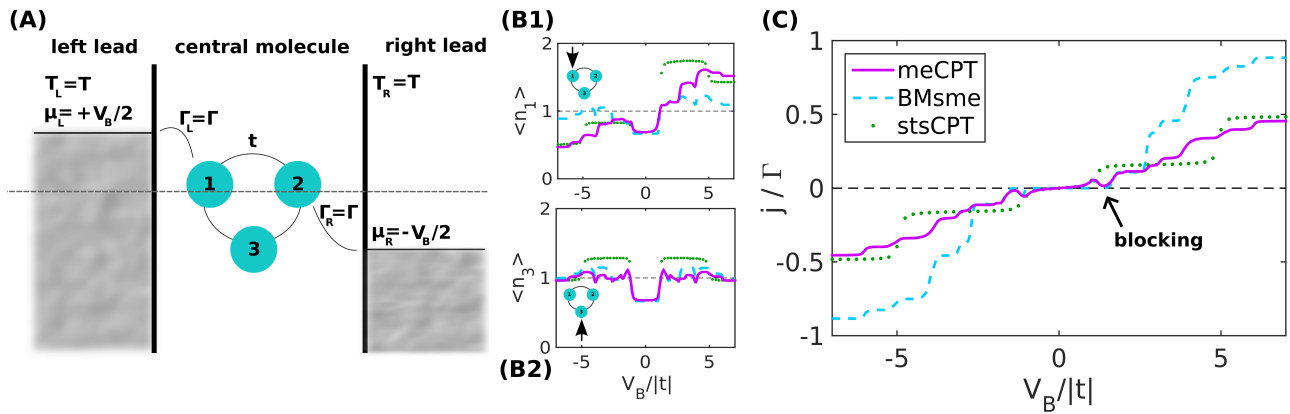}
\caption{(Color online) \emph{Triple quantum dot} \textbf{(A)} Schematic representation, see \se~\ref{ssec:QiBlocking}. System Hamiltonian as defined in \eq{eq:HS_tripledot}. Site 1 couples to the left lead and site 2 to the right one, both with $\Gamma_{\rm{L/R}}=\frac{\Gamma}{2}$. The leads are held at the same temperature $T_{\rm{L/R}}=T$ and the chemical potentials  $\mu_{\rm{L/R}}=\pm\frac{V_B}{2}$ are shifted by the bias voltage.
\textbf{(B)} Local charge density $\langle n_i\rangle$ as a function of bias voltage $V_B$. The results are obtained by meCPT, BMsme and stsCPT, see color code of panel (C). \textbf{(C)} Total current $j=\sum_\sigma \langle j_{L1 \sigma}\rangle$ into the system at site 1 as a function of bias voltage $V_B$. Results, shown in panels (B,C), are for $U=2\,|t|$, $T\approx0.02\,|t|$, $\Gamma=0.1\,|t|$ and $V_G=0$, corresponding to line (X) in \fig{fig:fig5}.}
\label{fig:fig4}
\end{figure*}
\begin{figure}
\myincludegraphics[width=0.49\textwidth]{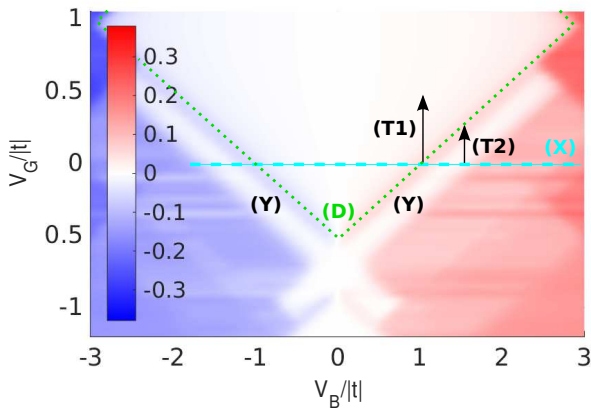}
\caption{(Color online) \emph{Triple quantum dot: stability diagram.} 
Total current entering the system as a function of bias voltage $V_B$ and gate voltage $V_G$, obtained within meCPT. The blocking region is indicated by (Y), the Coulomb diamond by (D). The two arrows (T1) and (T2) mark two device operation modes as discussed in the text. All results are for $U=2\,|t|$, $T=0.02\,|t|$ and $\Gamma=0.1\,|t|$. Dashed line (X) for $V_{G}=0$ marks the parameter region depicted in \fig{fig:fig4} (C).}
\label{fig:fig5}
\end{figure}

\begin{figure*}
\myincludegraphics[width=0.99\textwidth]{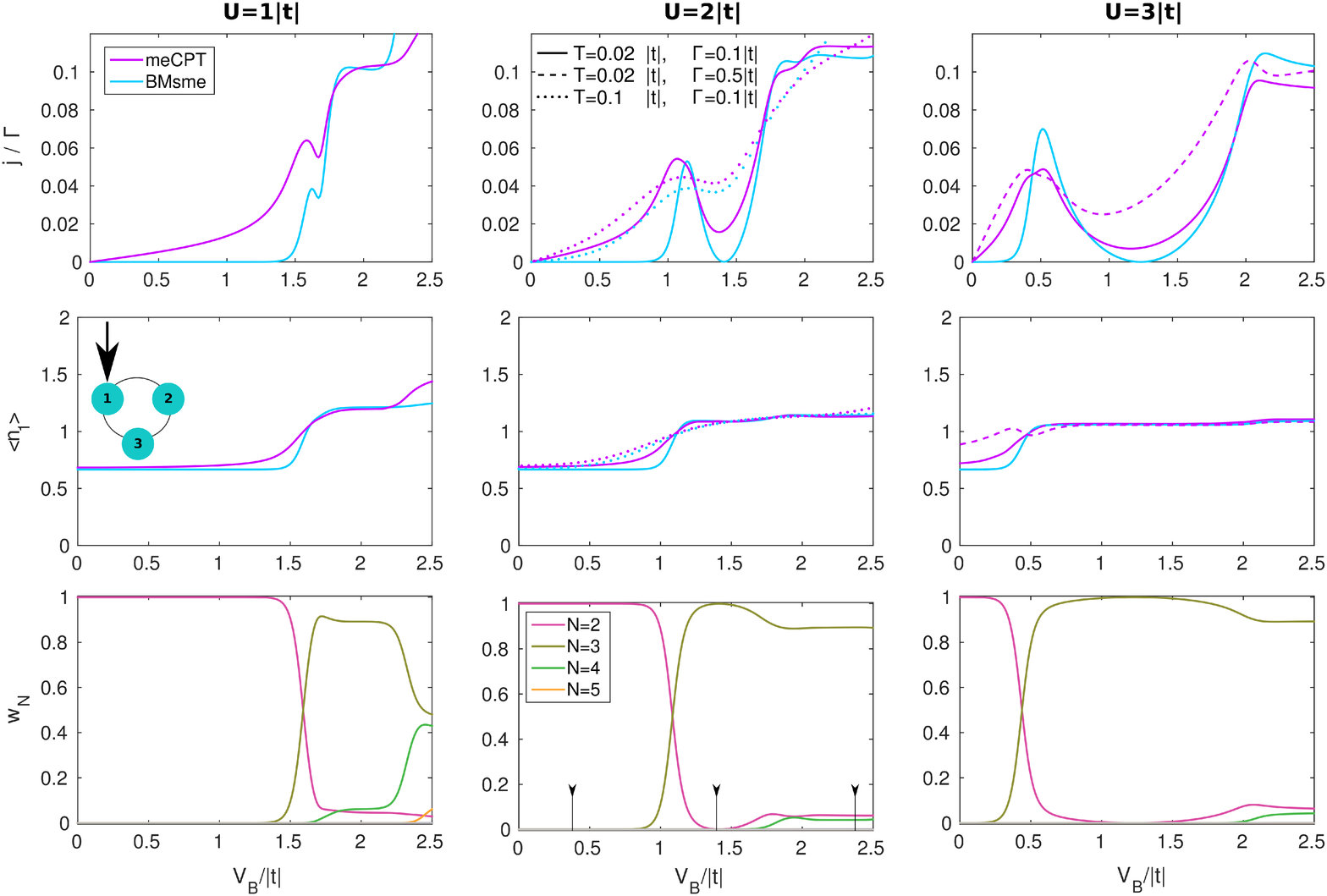}
\caption{(Color online) \emph{Triple quantum dot:} Dependence of the current blocking on the interaction strength $U$. \textbf{(Top row)} Total current $j$ as a function of bias voltage $V_B$. \textbf{(Middle row)} Charge density $\langle n_1\rangle$ at site $1$. The color code of the top row is valid. \textbf{(Bottom row)} Summed diagonal elements of the density matrix $w_N = \sum_{a\in N} \rho^S_{aa}$ per particle number $N$. The black markers in the mid panel ($U=2\,|t|$) indicate for which $V_B$ detailed results are given in \fig{fig:fig8}. Solid lines in all panels are for $T=0.02\,|t|$, $\Gamma=0.1\,|t|$ and $V_G=0$. Results for $T=0.1\,|t|$ are depicted in the central panels by dotted lines and those for $\Gamma=0.5\,|t|$ in the right panels by dashed lines.}
\label{fig:fig6}
\end{figure*}

\begin{figure*}
\myincludegraphics[width=0.99\textwidth]{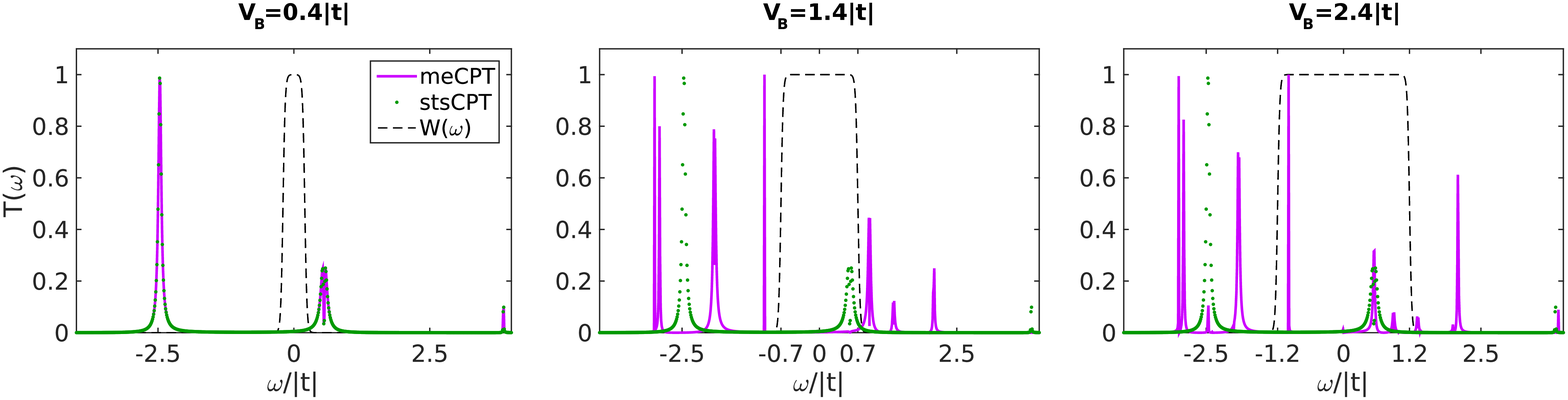}
\caption{(Color online) \emph{Triple quantum dot:} 
Dynamic transmission function $\mathcal{T}(\omega)$, \eq{eq:T}, as obtained by meCPT and stsCPT. Same parameters as in \fig{fig:fig6} (bottom mid) at the three indicated bias voltages: $V_B=0.4\,|t|$ (left), $V_B=1.4\,|t|$ (middle) and $V_B=2.4\,|t|$ (right). The temperature broadened transport window $W(\omega)$, \eq{eq:Wtrans}, is depicted as a dashed black line.}
\label{fig:fig8}
\end{figure*}

%% INTRO
In this section we discuss a more elaborate model system: a triple quantum dot ring junction which features negative differential conductance (NDC) based on electron-electron interaction effects mediated by quantum interference due to degenerate states as outlined in detail in \tcite{do.be.09, do.be.10}.
%% DEFINE MODEL
\Fig{fig:fig4} (A) depicts the triple quantum dot ring junction,
described by the following Hubbard Hamiltonian~\cite{hubb.63}
\begin{align}\label{eq:HS_tripledot}
 \hat{\mathcal{H}}^S&=\sum\limits_{i=1}^3 U 
 \big(\hat{n}^f_{i\uparrow}-\frac{1}{2}\big) 
 \big(\hat{n}^f_{i\downarrow}-\frac{1}{2}\big) 
 +V_{G}\;\sum\limits_{i=1}^3 \sum\limits_{\sigma} \hat{n}^f_{i\sigma}\nonumber\\
&+t\sum\limits_{\langle ij\rangle}\sum\limits_{\sigma} f^\dag_{i\sigma} f^\nag_{j\sigma}\,\mbox{.}
\end{align}
In addition to the model parameters described in \se~\ref{ssec:MBlocking}, a nearest-neighbour $\langle ij\rangle$ hopping $t$ is present.
% DEFINE ENVIRONMENT
The environment, \eq{eq:He} and coupling, \eq{eq:Hc} are now both symmetric in spin. Moreover, we use $\mu_L=-\mu_R=\frac{V_B}{2}$, $T=T_L=T_R$ and $\Gamma_L=\Gamma_R=\frac{\Gamma}{2}=\pi|t'|^2\frac{1}{2 D}$.

%% HOW TO REALIZE
Such a junction can be experimentally realized: i) Via local anodic oxidation (LAO) on a GaAs/AlGaAs heterostructure~\cite{ro.ha.08} which enables tunable few electron control.~\cite{ga.ka.09} ii) In a graphene nano structure.~\cite{mo.dr.09, er.we.09} Experimentally the stability diagram has been explored~\cite{ga.st.06} alongside characterisation and transport measurements.~\cite{ro.ha.08, ga.sa.09, au.pa.10} The negative differential conductance has been observed in a device aimed as a quantum rectifier.~\cite{vi.we.04} Theoretically the study of the nonequilibrium behaviour of such a device has become an active field recently.~\cite{de.sh.08, go.zh.08, do.be.09, ko.to.09, sh.de.09, po.em.09, emar.07, bu.sa.10, do.be.10} 

%% DISCUSS PHYSICS
We investigate transport properties for values of the parameters such that the junction is in a single electron transistor (SET) regime,~\cite{ko.ma.97} i.e. temperatures above the Kondo temperature.~\cite{hews.97} In this regime we expect an interaction  induced - quantum interference mediated blocking as discussed in \tcite{do.be.09, do.be.10}. The rotational symmetry ensures degenerate eigenstates labelled by a quantum number of angular momentum. In situations where these degenerate states participate in the transport they provide two equivalent pathways through the system and lead to quantum interference.~\cite{do.be.09} The blocking sets in at a bias voltage, where the degenerate states start to participate in the transport. It then becomes possible that a superposition is selected which forms one state with a node at the right lead. In the long time limit this state will be fully occupied while the other one will be empty due to Coulomb repulsion, for reasons very similar to the ones discussed in the 
previous section.~\cite{be.da.08, da.be.09}

%% PRESENT RESULTS
The steady-state charge distribution and current-voltage
characteristics of the interacting triple quantum dot are presented in
\fig{fig:fig4} (B, C) in a wide bias voltage window. The current, depicted in panel (C), in general increases in a stepwise manner and is fully antisymmetric with respect to the bias voltage direction. A blocking effect occurs at $V_B\approx1.5\,|t|$ as can be observed in the BMsme and meCPT data. The previous version of stsCPT based on the pure zero temperature ground state $\ket{\Psi_0}_S$ misses this region of NDC. In contrast to the simpler model presented in the previous section, a self-consistent mean-field solution does not capture the
blocking effects correctly in this more elaborate system. The BMsme
solution shows many more steps in the current than the stsCPT one, 
which is due to transitions in the reference state $\hat{\rho}^S$ of the central region. The meCPT results in general follow these finer steps, correcting their width to incorporate also lead induced broadening effects in addition to the pure temperature broadening. As can be seen in panel (B1), meCPT predicts a large charge increase at the site connected to the high bias lead. Note that the charge density at site $2$, which is connected to the right lead is simply: $\langle n_2\rangle(V_B)=\langle n_1\rangle(-V_B)$. The charge density at site 3 is symmetric with respect to the bias voltage origin. 

Next we study the impact of a gate voltage on the
blocking. Results obtained by meCPT are depicted as stability
diagram in \fig{fig:fig5}. Upon increasing $|V_G|$, the onset of the
blocking shifts linearly to higher $V_B$ (Y). We find a Coulomb
diamond for $2V_G \gtrapprox V_B-|t|$ (D). Upon increasing the bias voltage out of the Coulomb diamond, see e.g. line (X), a current sets in but is promptly hindered by the blocking so that the current diminishes after a hump of width $\propto
\text{max}(T,\Gamma)$. Interestingly this device could be operated as
a transistor in two fundamentally different modes. In mode (T1), at a
source-drain voltage of $\approx |t|$ the current is on for a gate voltage of $V_G=0$ and off for $V_G\approx0.5\,|t|$ due to the Coulomb blockade. In mode (T2), at a source-drain voltage of $\approx 1.5\,|t|$ the current is off for a gate voltage of $V_G=0$ due to quantum interference mediated blocking and on for $V_G=0.25\,|t|$.

Next we discuss the current characteristics in the vicinity of the blocking in more detail, as well as the impact of the interaction strength $U$. The first row of \Fig{fig:fig6} shows the total current through the device for different values of $U$. The blocking region shifts to lower bias voltages with increasing $U$. As discussed earlier, structures in the BMsme results are only broadened by temperature effects in the steady-state density (compare e.g. the width of the structures in the local density in the second row of \Fig{fig:fig6}), while meCPT additionally takes into account the finite life time of the quasi particles due to the coupling to the leads, given by $1/\Gamma$. This can be seen by solving \eq{eq:CPT} for the local  Green's function at device sites. Especially for higher lead broadening $\Gamma$ this gives rise to significant differences in the meCPT results compared to the BMsme data. 
From the bottom row of \Fig{fig:fig6} we see that, before the blocking
regime is entered, the steady-state changes from a pure $N=2$ state to
a mixed $N=2\,/\,N=3$ state at the hump in the current. Obviously,
blocking arises because the system reaches a pure $N=3$ state for
$U=2\,|t|$ and $U=3\,|t|$ at $V_B\approx1.4\,|t|$. For $U=|t|$ 
 the current is only partially blocked, because the contribution of the $N=2$ state is not fully suppressed. For all $U$-values, however, we find NDC.
As far as the meCPT current is concerned, the complete blocking at
higher interaction strengths, predicted by BMsme, is reduced to a
partial blocking due to the lead induced broadening effects in meCPT. Although
$\rho^S_{ab}$ changes significantly twice in the blocking region (for $U=2$ and $U=3$), the charge density $\langle n_i\rangle$ just increases once from $\langle n_1\rangle\approx0.75$ to $\langle n_1\rangle\approx1$.

Details of the steady-state dynamics are provided in
\fig{fig:fig8}. Before the blocking region is entered ($V_B=0.4\,|t|$) the
system is in a pure state with $N=2$, which corresponds to 
the zero temperature ground state $\ket{\Psi_0}_S$ in the $N=2$ sector. Here the transmission function $\mathcal{T}(\omega)$, \eq{eq:T}, of meCPT agrees with the one of stsCPT. A small current is obtained due to the $N=2\rightarrow3$ excitation at $\omega\approx0.55\,|t|$. 
Increasing the bias voltage has no influence on the reference state in
stsCPT, which therefore remains in the $N=2$ particle sector. Consequently, the transmission function in stsCPT does not change. Only the transport window increases linearly with increasing $V_{B}$. For $V_{B}=1.4\,|t|$ it includes the peak at $\approx 0.7|t|$ and results in a significant increase in the current obtained in stsCPT (see stsCPT result in \fig{fig:fig4}). This is in stark contrast to the BMsme current, depicted in \fig{fig:fig6}, which exhibits perfect blocking for  $V_B=1.4\,|t|$. The reason for the current-blocking is that only two states, both in the $N=3$ sector and doubly degenerate, have significant weight in $\rho^S_{ab}$. The meCPT solution is based on the modified density matrix and therefore the current is diminished, since the next possible excitation is at $\omega\approx0.9\,|t|$ ($N=2\rightarrow3$), which is outside the transport window $W(\omega)\approx (-0.7|t|,0.7|t|)$, \eq{eq:Wtrans}.
Due to the lead induced broadening of $\mathcal{T}(\omega)$ and the
temperature induced broadening of the transport window, the current is
however only partially blocked. For $V_B=2.4\,|t|$ this excitation
falls into the transport window and the current is no longer blocked.
In this case, the state $\rho^S_{ab}$ is a mixture of $N=2,3,4$. The dominant excitation responsible for this current is again the ground state excitation at $\omega\approx0.55\,|t|$ from $N=2\rightarrow3$. This is why in this regime the stsCPT current, based on the pure two particle state is again similar to the meCPT current.

Our results on the Qme level have been checked with those presented by Begemann \etal{} in \tcite{be.da.08} and Darau \etal{} in \tcite{da.be.09} for a six orbital ring which shows similar blocking effects. Different types of blocking effects in various parameter regimes have been discussed in detail in a Qme framework also for the three orbital ring by Donarini \etal{} in  \tcite{do.be.09, do.be.10}.

\subsubsection*{Quasi-degenerate states}\label{ssec:Vg3}
%% INTRO, DEF MODEL
Next we study the reliability of the secular approximation in the case
of quasi degeneracy of the isolated energies of the system and benchmark its applicability to create a reference state for meCPT. To this end we apply a second gate voltage that couples only to the third orbital, see \fig{fig:fig4} (left), and leads to an additional  term $ V_{G,3}\, \hat{n}^f_{3}$ in the system Hamiltonian. This lifts the degeneracy of states present at $V_{G,3}=0$ and therefore requires a treatment within the BMme, see \tcite{da.be.09}.

\begin{figure}
\myincludegraphics[width=0.4\textwidth]{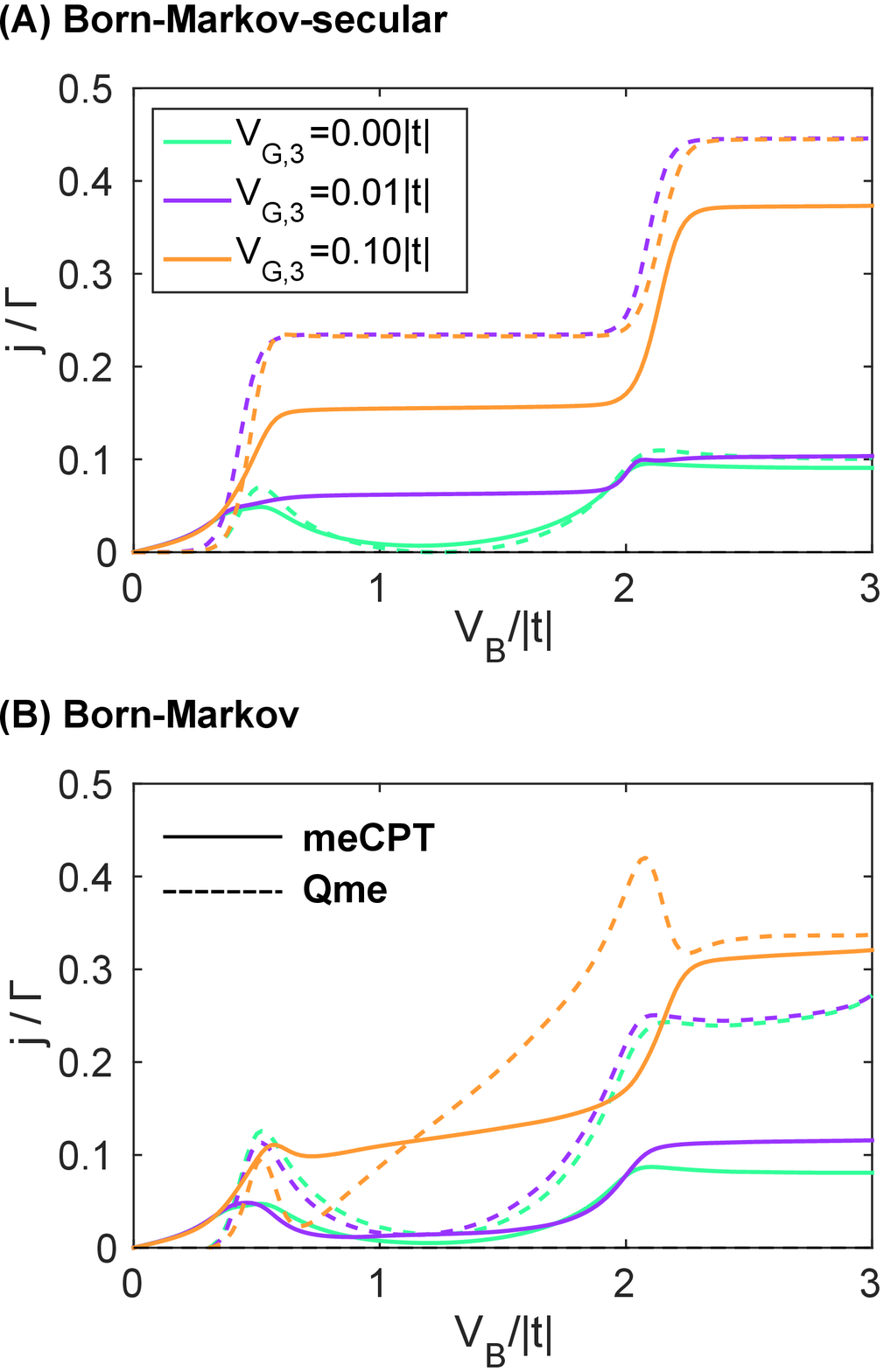}
\caption{(Color online) \emph{Triple quantum dot:} Effects of lifting degeneracies in the system energies by a third gate voltage.
Total current $j$ as a function of bias voltage $V_B$, for three different gate voltages $V_{G,3}$ applied to site $3$.
%$V_{G,3}=0\,|t|$, $V_{G,3}=0.01\,|t|$ and $V_{G,3}=0.1\,|t|$. 
Results based on the Born-Markov-secular approximation are compared with those of the Born-Markov approximation. Solid/dashed lines indicate the meCPT/BM(s)me result. All results are for $U=3\,|t|$, $T=0.02\,|t|$ and $\Gamma=0.1\,|t|$.}
\label{fig:fig7}
\end{figure}

%% RESULTS
In the following we discuss the same parameter regime as above. In
\fig{fig:fig7} we present results obtained using meCPT (solid lines) and Qme results (dashed lines) for the BMsme (A) and for the BMme (B). The meCPT results of each panel are obtained using the respective Qme. In the BMsme data a very small $|V_{G,3}|$ has a drastic effect on the current-voltage characteristics. The blocking present at $V_{G,3}=0$ is immediately lifted by very small $|V_{G,3}|$ and the current jumps to a plateau. For larger $|V_{G,3}|$ the current stays on this plateau until further transport channels open up. This ''jump`` at small $|V_{G,3}|$ arises due to the improper treatment of quasi degeneracies in BMsme. MeCPT results based on BMsme show a smooth change of the current-voltage characteristics. BMme on the other hand correctly accounts for the coupling of the quasi-degenerate states and also exhibits a smooth dependence on $V_{G,3}$. For meCPT based on BMme we find qualitative similar results to meCPT based on BMsme, which emphasizes the robustness of the meCPT 
results in general. From 
this it is apparent that meCPT is capable of repairing the decoupling of quasi-degenerate states in the BMsme to some degree. However, to study blocking effects at quasi degenerate points it is of advantage to make use of the BMme in meCPT.

%% ISSUES BMME
As discussed below in \se~\ref{ssec:cl}, the BMme is not of Lindblad
form and does not necessarily result in a positive definite reduced
many-body density matrix $\rho^S_{a b}$ in general. Using a not proper
density matrix in \eq{eq:grl} may result in non-causal Green's functions
when the steady-state $\rho^S_{a b}$ is obtained from the BMme. This
can be avoided by using a modified reference state $\rho^S_{a b}
\rightarrow \rho^S_{a b}\Theta(\Delta - |\omega_a-\omega_b|)$, with
$\Theta(x)$ the Heaviside step function and $\Delta$ a small quantity, being
e.g. $\approx 10^{-6}$, in \eq{eq:grl}, which renders the Green's functions
causal. This is somewhat an ad-hoc approximation and should be seen simply as
a way to explore the effects of continuously breaking degeneracy in
the problem.

\subsection{Current conservation}\label{ssec:cl}
Finally we comment on conservation laws in meCPT. Within BMsme and BMme the current conservation (continuity equation) is always maximally violated in a sense that the current within the system is zero. This is due to the zeroth order $\hat{\rho}^S$ as discussed in \app~\ref{app:observables2}. In BMsme the inflow from the left lead into the system however always equals the outflow from the system to the right lead.  Without the secular approximation the quantum master equation (BMme) is not of Lindblad form and the final many-body density matrix is not guaranteed to be positive definite.~\cite{whit.08,ya.sh.00} This in turn can lead to slightly negative currents in regions where they are required to be positive by the direction of the bias voltage~\cite{scha.12}. Furthermore, the inflow can be slightly different from the outflow. 

In the noninteracting case, meCPT fully repairs the violation of the continuity equation present in the reference state. For increasing interaction strength, the violation of the continuity equation typically grows also in meCPT. In particular, the overall symmetry of the current stays intact (in our case, inflow equals outflow), while the current on bonds between interacting sites does not exactly match the current between noninteracting sites. This typically small violation of the continuity equation can be attributed to the violation of Ward identities~\cite{ward.50, en.sc.63} in the non-conserving approximation scheme of CPT.~\cite{ba.ka.61, baym.62}

\section{Summary and Conclusions}\label{sec:conclusion}
We improved steady-state cluster perturbation theory with an appropriate, consistent reference state. This reference state is obtained by the reduced many-body density matrix in the steady-state obtained from a quantum master equation. The resulting hybrid method inherits beneficial aspects of steady-state cluster perturbation theory as well as from the quantum master equation.

We benchmarked the new method on two experimentally realizable systems: a quantum diode and a triple quantum dot ring, which both feature negative differential conductance and interaction induced current blocking effects. meCPT is able to improve the bare quantum master equation results by a correct inclusion of lead induced level-broadening effects, and the correct noninteracting limit. In contrast to previous realizations of the steady-state cluster perturbation theory, meCPT is able to correctly predict interaction induced current blocking effects. 
It is well known that the secular approximation (BMsme) is not
applicable to quasi degenerate problems, which is corroborated by our
results for the steady-state current. However, meCPT based on the
BMsme density, is able to repair most of the shortcomings of BMsme. 
The results are very close to those obtained by meCPT based on the density of BMme, where the quasi-degenerate states are treated consistently.

The computational effort of meCPT beyond that of the bare quantum master equation scales with the number of significant entries in the reference state density matrix but is typically small. In the presented formulation the new method is flexible and fast and therefore well suited to study nano structures, molecular junctions or heterostructures also starting from an \textit{ab-inito} calculation.~\cite{ry.do.13}

\begin{acknowledgments}
The authors acknowledge fruitful discussion with A. Rosch. This work was partly supported by the Austrian Science Fund (FWF) Grants No. P24081 and P26508 as well as SFB-ViCoM projects F04103 and F04104 and NaWi Graz. MN, GD and AD thank the Forschungszentrum J{\"u}lich, in particular the autumn school on correlated electrons, for hospitality and support.
\end{acknowledgments}

\appendix

\section{Born-Markov and Pauli master equation}\label{app:BMdetails}
Here we provide the detailed expressions for the coefficients in the BMme and BMsme of ~\eq{eq:BMsme} and discuss the equations governing the time evolution into the steady-state.

%% DEFINITIONS
The Lamb-shift Hamiltonian is defined as $\hat{\mathcal{H}}^{LS} = \sum\limits_{ab} \Lambda_{ab}\ket{a}\bra{b}$, with
\begin{align}
\label{eq:Lambda}\Lambda_{ab} &= \frac{1}{2i}\sum\limits_{\alpha\beta}\sum\limits_c \lambda_{\alpha\beta}(\omega_{bc},\omega_{ac})\bra{c}\hat{S}_\beta\ket{b} \bra{c}\hat{S}_\alpha\ket{a}^*\,\mbox{.}
\end{align}
Note that $[\hat{\mathcal{H}}^{LS},\hat{\mathcal{H}}^{S}]_- = 0$. In the secular approximation (BMsme) one can replace $\lambda_{\alpha\beta}(\omega_{bc},\omega_{ac}) \rightarrow \lambda_{\alpha\beta}(\omega_b-\omega_c) \delta_{\omega_b,\omega_a}$. The expressions for the BMme and BMsme \eq{eq:BMsme} are valid if $\left[\hat{\mathcal{H}}^E,{\hat{\rho}}^E\right]_-=0\;\mbox{ and } \mytr\left\{\hat{E}_\alpha {\hat{\rho}}^E\right\}=0$. The environment functions $\xi_{\alpha\beta}$ and $\lambda_{\alpha\beta}$ in \eq{eq:Lambda} and \eq{eq:Xi} are determined by the time dependent environment correlation functions
\begin{align}
\label{eq:Ctau} C_{\alpha\beta}(\tau) &=\text{tr}\left\{\hat{E}_\alpha(\tau) \hat{E}_\beta {\hat{\rho}}_E\right\}\,\mbox{,}
\end{align}
where the Heisenberg time evolution in the environment operators is $\hat{E}_\alpha(\tau) = e^{+i \hat{\mathcal{H}}^E\tau}\hat{E}_\alpha e^{-i  \hat{\mathcal{H}}^E\tau}$.

For the BMme, $\xi_{\alpha\beta}$ and $\lambda_{\alpha\beta}$ are given by a sum of complex Laplace transforms
\begin{align}
\label{eq:xi_nosec} \xi_{\alpha\beta}(\omega_1, \omega_2) &= \int_0^\infty d\tau\,C_{\alpha\beta}(\tau)e^{+i \omega_1 \tau} + \int_{-\infty}^0 d\tau\,C_{\alpha\beta}(\tau)e^{+i \omega_2 \tau}\,\mbox{,}\\
\label{eq:lambda_nosec} \lambda_{\alpha\beta}(\omega_1, \omega_2) &= \int_0^\infty d\tau\,C_{\alpha\beta}(\tau)e^{+i \omega_1 \tau} - \int_{-\infty}^0 d\tau\,C_{\alpha\beta}(\tau)e^{+i \omega_2 \tau}\,\mbox{,}
\end{align}
whereas for the BMsme ($\omega_1=\omega_2$) the expressions simplify to the full even and odd Fourier transforms~\cite{scha.12}
\begin{align}
\label{eq:xi} \xi_{\alpha\beta}(\omega) &= \int\limits_{-\infty}^\infty d\tau\,C_{\alpha\beta}(\tau)e^{+i \omega \tau}\,\mbox{,}\\
\label{eq:lambda} \lambda_{\alpha\beta}(\omega) &= \int\limits_{-\infty}^\infty d\tau\,\text{sign}(\tau) C_{\alpha\beta}
(\tau)e^{+i \omega \tau} = \frac{i}{\pi}\intP{-\infty}{\infty}\,d\omega' 
\frac{\xi_{\alpha\beta}(\omega')}{\omega-\omega'}\,\mbox{.}
\end{align}

%% EQUATIONS FOR TIME EVOLVE
The coupled equations for the real time evolution of the components of the reduced system many-body density matrix $\rho^{S}_{ab}=\bra{a}\hat{\rho}^{S}\ket{b}$ according to the BMsme read
\begin{align}
\dot{\rho}^{S}_{ab}(\tau)&= i(\omega_b-\omega_a)\rho^{S}_{ab}(\tau)\\
\nonumber&+i\sum\limits_{c}\bigg(\rho^{S}_{ac}(\tau)\Lambda_{cb}
-\Lambda_{ac}\rho^{S}_{cb}(\tau)\bigg)\\
\nonumber&+\sum\limits_{cd}\Bigg( \Xi_{ac,bd} \rho^{S}_{cd}(\tau)
- \frac{1}{2} \Xi_{cd,ca}\rho^{S}_{db}(\tau) \\
\nonumber&- \frac{1}{2}\Xi_{cb,cd}\rho^{S}_{ad}(\tau)\Bigg)\,\mbox{.}
\end{align}

The equations simplify further for system Hamiltonians $\hat{\mathcal{H}}^S$ with non-degenerate eigenenergies $\omega_a$. Then the diagonal components $\phi_{a}=\rho^S_{aa}$ decouple from the off-diagonals and one recovers the Pauli master equation for classical probabilities
\begin{align}
\label{eq:psi}
\dot{\phi}_{a}(\tau)&= \sum\limits_{c} \bigg(\Xi_{ac}\phi_{c}(\tau)- \Xi_{ca}\phi_{a}(\tau)\bigg)\,\mbox{,}
\end{align}
with simplified coefficients 
\begin{align*}
\Xi_{ab}&:= \Xi_{ab,ab} = \sum\limits_{\alpha\beta} \xi_{\alpha\beta}(\omega_b-\omega_a)\bra{a}\hat{S}_\beta\ket{b} \bra{a}\hat{S}_\alpha\ket{b}^*\,\mbox{.}
\end{align*}
In this case the dynamics of the decoupled off-diagonal terms  ($a\neq b$) is given by
\begin{align*}
\dot{\rho}^{S}_{ab}(\tau) &= \Bigg(i (\omega_b + \Lambda_b - \omega_a - \Lambda_a)\\
&- \frac{1}{2}\sum\limits_{c}\bigg( \Xi_{ca} + \Xi_{cb}\bigg)\Bigg)\rho^{S}_{ab}(\tau)\,\mbox{,}
\end{align*}
where the simplified Lamb shift terms are
\begin{align*}
\Lambda_{a} &:= \Lambda_{aa} = \frac{1}{2i}\sum\limits_{\alpha\beta}\sum\limits_c \lambda_{\alpha\beta}(\omega_a-\omega_c) \bra{c}\hat{S}_\beta\ket{a} \bra{c}\hat{S}_\alpha\ket{a}^*\,\mbox{.}
\end{align*}

\section{Hermitian tensor product form of the coupling Hamiltonian}\label{app:tensorProduct}
For the BMsme (see \se~\ref{ssec:BMsme}) it is necessary to bring the fermionic system-environment coupling Hamiltonian, \eq{eq:Hc} to a hermitian tensor product form, which requires $[\hat{S}_\alpha,\hat{E}_\alpha]_-=0$. For the fermionic operators in \eq{eq:Hc} we however have $[f_{i\sigma}^\dag,c_{\lambda k \sigma}]_-= 2 f_{i\sigma}^\dag c_{\lambda k \sigma}$. A solution is provided in \tcite{scha.14} by performing a Jordan-Wigner transformation~\cite{jo.wi.28} on the system and environment operators
\begin{align*}
f_{i\sigma} &= \prod\limits_\sigma \left(\xi_{1}^z\otimes\ldots\otimes\xi_{i-1}^z \xi_{i}^- \uu_{i+1}\otimes\ldots\otimes\uu_{L_S}\right)_{S,\sigma}\\
&\otimes\prod\limits_\lambda \left(\uu_{1}\otimes\ldots\otimes\uu_{L_E}\right)_{E,\lambda\sigma}\,\mbox{,}\\
c_{\lambda j \sigma} &= \prod\limits_\sigma \left(\xi_{1}^z\otimes\ldots\otimes\xi_{L_S}^z\right)_{S,\sigma}\otimes\\
&\prod\limits_\lambda \left(\eta_{1}^z\otimes\ldots\otimes\eta_{j-1}^z \eta_{j }^- \uu_{j+1}\otimes\ldots\otimes\uu_{L_E}\right)_{E,\lambda\sigma}\,\mbox{,}
\end{align*}
where $\xi_{i}$ and $\eta_{j}$ denote local spin-$\frac{1}{2}$ degrees of freedom at the system and environment sites respectively and the overall ordering of operators is important. $L_S/L_E$ denote the size of the system / environment. Reordering \eq{eq:Hc} we find $\hat{\mathcal{H}}^{SE}_\lambda =\sum\limits_{ij\sigma}t'_{\lambda i j \sigma} f_{i\sigma}^\dag c_{\lambda j \sigma}^\nag - t'^*_{\lambda i j \sigma} f_{i\sigma}^\nag c_{\lambda j \sigma}^\dag$, where the minus sign arises due to the fermionic anti-commutator. Plugging in the Jordan-Wigner transformed operators leads to
\begin{align*}
 \hat{\mathcal{H}}^{SE}_\lambda&=\sum\limits_{ij\sigma}\bigg(t'_{\lambda i j \sigma}\bigg[\xi_i^+\otimes\\
&[-\xi_{i+1}^z\otimes\ldots\otimes\xi_{L_S}^z\otimes\eta_{1}^z\otimes\ldots\otimes\eta_{j-1}^z]\otimes
\eta_j^-\bigg]_{\sigma\lambda}+t'^*_{\lambda i j \sigma}\\
&\bigg[\xi_i^-\otimes[-\xi_{i+1}^z\otimes\ldots\otimes\xi_{L_S}^z\otimes
\eta_{1}^z\otimes\ldots\otimes\eta_{j-1}^z]\otimes
\eta_j^+\bigg]_{\sigma\lambda}\bigg)\\
&=\sum\limits_{i}\left(\bar{f}_i^\dag\otimes\bar{c}_i + \bar{f}_i\otimes\bar{c}_i^\dag \right)\,\mbox{,}
\end{align*}
where in the last line we have defined new operators
\begin{align*}
 \bar{f}_{i\sigma}&=\left[\xi_i^-\otimes[-\xi_{i+1}^z\otimes\ldots\otimes\xi_{L_S}^z]\right]_\sigma\,\mbox{,}\\
\bar{f}_{i\sigma}^\dag&=\left[[-\xi_{i+1}^z\otimes\ldots\otimes\xi_{L_S}^z]\otimes\xi_i^+\right]_\sigma\,\mbox{,}\\
\bar{c}_{\lambda i\sigma}&=\sum\limits_j t'_{\lambda ij \sigma}\left[[\eta_{1}^z\otimes\ldots\otimes\eta_{j-1}^z]\otimes\eta_j^-\right]_{\lambda\sigma}\,\mbox{,}\\
\bar{c}_{\lambda i\sigma}^\dag&=\sum\limits_j t'^*_{\lambda ij \sigma}\left[\eta_j^+\otimes[\eta_{1}^z\otimes\ldots\otimes\eta_{j-1}^z]\right]_{\lambda\sigma}\,\mbox{.}
\end{align*}
Note that the phase operator $\hat{P}_{i(j\lambda)\sigma}=\left[-\xi_{i+1}^z\otimes\ldots\otimes\xi_{L_S}^z\otimes[\eta_{1}^z\otimes\ldots\otimes\eta_{j-1}^z]_\lambda\right]_\sigma = (-1)^{1 + \sum\limits_{\lambda'} \sum\limits_{m=i+1}^{L_S}\hat{n}_m + \hat{N}_{j\lambda'}}$ counts the particles between system site $i$ and environment site $j$ for spin $\sigma$ depending on the ordering of the environments $\lambda$. It is straight forward to show that the bar operators fulfil fermionic anti-commutation rules. Furthermore $[\bar{f}_{i\sigma},\bar{c}_{\lambda i\sigma}]_-=0$, which allows us to write the coupling Hamiltonian in a tensor product form. Note that in general  $[\bar{f}_{i\sigma},\bar{c}_{\lambda' j\sigma}]_-\neq 0$ for $i\neq j$ which is however not relevant for the coupling Hamiltonian where only the same $i$ couple.

The new operators in hermitian form are given in \eq{eq:couplingOperators} by replacing $c\rightarrow \bar{c}$ and $f\rightarrow \bar{f}$. Next we show, by examining the BMsme, that in most cases the additional phase operator in $\bar{c}$ drops out of the calculations and we are even allowed to use the original $f$ and $c$ operators instead of the barred ones. The operators $\bar{c}$ only enter the equations in the environment correlation functions $C_{\alpha\beta}(\tau)$ as defined in \eq{eq:Ctau}. Plugging in the barred operators we obtain for normal systems which preserve particle number
\begin{align*}
 C_{\alpha\beta}(\tau) &\propto \mytr\left\{e^{+i \hat{\mathcal{H}}^E\tau} f_{\lambda j\sigma}^\dag  e^{-i  \hat{\mathcal{H}}^E\tau}\hat{P}_{i(j\lambda)\sigma}^2 c_{\lambda j\sigma}{\hat{\rho}}^E\right\}\,\mbox{,}
\end{align*}
with $\hat{P}_{ij}^2=\uu$, where we required that $[\hat{\mathcal{H}}^E,\hat{P}_{ij}]_- = 0$. The dropping out of the phase operators implies that for normal systems where the disconnected environments conserve particle number we can omit the Jordan-Wigner transformation and do all calculations as is with the original environment creation/annihilation operators in hermitian form.

\section{Bath correlation functions}\label{sssec:environmentCorrelators}
In the wide band limit, analytical expressions for the bath correlation functions are available in \tcite{be.da.08}. For arbitrary environment DOS, explicit evaluation of the environment correlation functions becomes convenient for hermitian couplings, \eq{eq:couplingOperators} as outlined in \app~\ref{app:tensorProduct}.~\cite{scha.12} Essentially the environment functions can all be obtained via integrals of the environment DOS $\rho(\omega)$. Care has to be taken when going to very low temperatures and solving the integrals with finite precision arithmetic to avoid underflow errors.

The time dependent environment correlation functions $C_{\alpha\beta}(\tau)$, \eq{eq:Ctau} become
\begin{align*}
 C_{11}(\tau) &= C_{22}(\tau) =\frac{1}{4 \pi}\sum\limits_{\lambda\sigma} \int\limits_{-\infty}^{\infty} d\nu\,
\Gamma_{\lambda\sigma}(\nu)\\
&\times\bigg(e^{-i\nu \tau} + 2i p_{\text{FD}}(\nu,T_\lambda,\mu_\lambda)\sin{(\nu\tau)}\bigg)\,\mbox{,}\\
C_{12}(\tau) &= -C_{21}(\tau) =\frac{i}{4 \pi}\sum\limits_{\lambda\sigma} \int\limits_{-\infty}^{\infty}d\nu\,
\Gamma_{\lambda\sigma}(\nu)\\
&\times\bigg(-e^{-i\nu \tau} + 2 p_{\text{FD}}(\nu,T_\lambda,\mu_\lambda)\cos{(\nu\tau)}\bigg)\,\mbox{,}
\end{align*}
where $C_{\alpha\beta}(\tau)=C_{\beta\alpha}^\dag(-\tau)$ and the coefficient 
\begin{align}
\Gamma_{\lambda\sigma}(\nu)=&2\pi |t'_{\lambda \sigma}|^2\sum\limits_k\delta(\nu-\omega_{\lambda k \sigma})\,
\mbox{,}
\label{eq:Gamma}
\end{align}
is proportional to the lead DOS.

For the BMsme, the respective full even Fourier transforms $\xi_{\alpha\beta}(\omega)$, \eq{eq:xi} we find
\begin{align*}
&\xi_{11}(\omega) = \xi_{22}(\omega) =\\ &\frac{1}{2}\sum\limits_{\lambda\sigma}
\Gamma_{\lambda\sigma}(-\omega) p_{\text{FD}}(-\omega,\beta_\lambda,\mu_\lambda)
+\Gamma_{\lambda\sigma}(\omega) \overline{p}_{\text{FD}}(\omega,T_\lambda,\mu_\lambda)\,\mbox{,}
\\
&\xi_{12}(\omega) = -\xi_{21}(\omega) =\\
& \frac{i}{2}\sum\limits_{\lambda\sigma}
\Gamma_{\lambda\sigma}(-\omega) p_{\text{FD}}(-\omega,\beta_\lambda,\mu_\lambda)
-\Gamma_{\lambda\sigma}(\omega) \overline{p}_{\text{FD}}(\omega,T_\lambda,\mu_\lambda)
\,\mbox{,}
\end{align*}
where $\overline{p}_{\text{FD}}(\omega,T,\mu) = 1-p_{\text{FD}}(\omega,T,\mu)$.

The odd Fourier transforms $\lambda_{\alpha\beta}(\omega)$, \eq{eq:lambda} are given by
\begin{align*}
&\lambda_{11}(\omega) = \lambda_{22}(\omega) =\\
&\frac{i}{2\pi}\sum\limits_{\lambda\sigma}
\intP{-\infty}{\infty}\,d\nu \Gamma_{\lambda\sigma}(\nu) \bigg(\frac{p_{\text{FD}}(\nu,\beta_\lambda,\mu_\lambda)}{\nu + \omega} - \frac{\overline{p}_{\text{FD}}(\nu,\beta_\lambda,\mu_\lambda)}{\nu-\omega}\bigg)\,\mbox{,}\\
&\lambda_{12}(\omega) = -\lambda_{21}(\omega) =\\
&-\frac{1}{2\pi}\sum\limits_{\lambda\sigma}\intP{-\infty}{\infty}\,d\nu \Gamma_{\lambda\sigma}(\nu) \bigg(\frac{
 p_{\text{FD}}(\nu,\beta_\lambda,\mu_\lambda)}{\nu+\omega}
+ \frac{ \overline{p}_{\text{FD}}(\nu,\beta_\lambda,\mu_\lambda)}{\nu-\omega}\bigg)\,\mbox{.}
\end{align*}

\section{Evaluation of steady-state observables}\label{app:observables}
\subsection{Steady-state cluster perturbation theory}\label{app:observables1}
Within meCPT single-particle observables are available by integration of $\widetilde{G}(\omega)$, \eq{eq:CPT}. Its easy to show that the single-particle density matrix $\kappa_{ij\sigma}= \frac{\delta_{ij}}{2} -\frac{i}{2}\int\limits_{-\infty}^\infty \frac{d\omega}{2 \pi}G^K_{ij\sigma}(\omega)$ can be expressed in terms of the retarded CPT Green's function
\begin{align*}
\kappa_{ij\sigma}&= \frac{\delta_{ij}}{2} -\frac{i}{2}\int\limits_{-\infty}^\infty \frac{d\omega}{2 \pi} \Bigg(G^R_{in\sigma}(\omega) P_{nj\sigma}(\omega) -P_{in\sigma}(\omega)(G^R_{jn\sigma}(\omega))^*\\
&+G^R_{in\sigma}(\omega)\left(\left[P_{\sigma}(\omega),M_{\sigma}\right]_-\right)_{nm}(G^R_{jm\sigma}(\omega))^*\Bigg)\,\mbox{,}
\end{align*}
where $M_{\sigma}$ is the inter-cluster perturbation defined in \eq{eq:CPT}. Here we use the Einstein summation convention, the last line holds within CPT and $P_{ij\sigma}(\omega) = \delta_{ij} (1-2 p_{\text{FD}}(\omega,T_i,\mu_{i\sigma}))$.

From the real part of the single-particle density-matrix we read off the site occupation $\langle n_i\rangle= \sum\limits_\sigma \kappa_{ii\sigma}$ the spin resolved occupations $\langle n_{i\sigma}\rangle = \kappa_{ii\sigma}$ and the magnetization $ \langle m_i \rangle  = \frac{1}{2}(\kappa_{ii\uparrow}-\kappa_{ii\downarrow})$.

The current $\langle j_{\langle ij\rangle}\rangle$ between nearest-neighbour sites $\langle ij \rangle $ is related to the imaginary part of $\kappa_{ij\sigma}$ and reads in symmetrized form
\begin{align*}
 \langle j_{\langle ij\rangle}\rangle&=\frac{e}{2\hbar}\left(h_{ij\sigma}\kappa_{ij\sigma}-h_{ji\sigma}\kappa_{ji\sigma}\right)\,\mbox{,}
\end{align*}
which is of Meir-Wingreen form~\cite{me.wi.92} and $h_{ij\sigma}$ is the single-particle Hamiltonian.

Equivalently, the transmission current between two environments $\lambda = 1,\,2$ can be evaluated in the Landauer-B\"uttiker form~\cite{ha.ja.96, datt.05, kn.ar.13}
\begin{align}
\langle j_{{1/2}}\rangle &= \frac{e}{\hbar}\int\limits_{-\infty}^\infty\frac{d\omega}{2 \pi} W(\omega) \text{tr}\left\{\mathcal{T}(\omega)\right\}\,\mbox{,}
\label{eq:jByT}
\end{align}
with the transport window
\begin{align}
 W(\omega) &= p_{\text{FD}}(\omega,T_{1},\mu_{1})-p_{\text{FD}}(\omega,T_{2},\mu_{2})\,\mbox{,}
 \label{eq:Wtrans}
\end{align}
and where the transmission function
\begin{align}
\mathcal{T}(\omega)&=G^R(\omega)\Gamma_{1}(\omega)\left(G^R(\omega)\right)^\dag\Gamma_{2}(\omega)\,\mbox{,}
\label{eq:T}
\end{align}
is given in terms of $G^R(\omega) = \left((g^R(\omega))^{-1}-(\widetilde{\Sigma}_{1}+\widetilde{\Sigma}_{2})\right)^{-1}$ with the lead broadening functions of lead $\lambda$ projected onto the system sites $i,j$ is $\widetilde{\Sigma}_{\lambda i j}=M_{i\lambda} g^R_{\lambda \lambda}M_{\lambda j}$ and $\Gamma_\lambda=-2\Im\text{m}\left(\widetilde{\Sigma}_{\lambda}\right)$, compare also \eq{eq:Gamma}.

\subsection{Born-Markov master equation}\label{app:observables2}
Within the Qme, basic single-particle observables are available in terms of the reduced system many-body density matrix $\hat{\rho}^S$. The single-particle density matrix $\kappa$ reads
\begin{align}
 \kappa_{ij\sigma} &= \text{tr}\bigg(f_{i\sigma}^\dag f_{j\sigma}^\nag \hat{\rho}^{S}\bigg)= \sum\limits_{ab} \bra{b} f_{i\sigma}^\dag f_{j\sigma}^\nag \ket{a}\rho^S_{ab}\,\mbox{,}
\label{eq:kappaBM}
\end{align}
where $a$ and $b$ denote eigenstates of the system Hamiltonian. Note that within the BMme/BMsme $\kappa_{ij\sigma}$ is purely real and therefore does predict zero current.

However, an expression for the current to reservoir $\lambda$ can be found by making use of the operator of total system charge $\hat{Q}$ and total system particle number $\hat{N}$, where $q$ denotes the charge of one carrier
\begin{align*}
 \sum\limits_\lambda j^\lambda(\tau) &= \frac{d}{d\tau}\langle\hat{Q}(\tau)\rangle=  q\,\text{tr}\left(\hat{N} \dot{\hat\rho}^S(\tau) \right)\,\mbox{.}
\end{align*}
Taking $\dot{\hat{\rho}}^S(\tau)$ from the Qme we obtain
\begin{align*}
j^\lambda &= q\sum\limits_{abc} \bigg(n_c  -\frac{1}{2} n_b-\frac{1}{2} n_a\bigg)\Xi^\lambda_{ca,cb} \bigg)\rho^S_{ab}\,\mbox{,}
\end{align*}
and for non-degenerate systems, in the Pauli limit we find from the BMsme
\begin{align*}
j^\lambda_{\text{non-deg}} &=q\sum\limits_{ab} \left(n_a - n_b\right)\Xi^\lambda_{ab} \phi_{b}\,\mbox{.}
\end{align*}

\bibliography{bmsme2necpt,footnotes}{}

\begin{thebibliography}{140}
\bibitem[{foo({\natexlab{a}})}]{footnote2}
\bibinfo{note}{Throughout this paper we use $\hbar= k_B\equiv 1$.}
\bibitem[{foo({\natexlab{c}})}]{footnote3}
\bibinfo{note}{Assuming that the interacting part is small enough to be treated
  exactly, otherwise it has to be decomposed into sub partitions again.}
\bibitem[{foo({\natexlab{d}})}]{footnote1}
\bibinfo{note}{We denote the largest energy scale or parameter of
  $\hat{\mathcal{H}}$ as $|\hat{\mathcal{H}}|$.}
\end{thebibliography}

\end{document}